\documentclass[conference]{IEEEtran}
\IEEEoverridecommandlockouts
\usepackage{sty/style}

\definecolor{crabred}{HTML}{BE1831}

\def\BibTeX{{\rm B\kern-.05em{\sc i\kern-.025em b}\kern-.08em
    T\kern-.1667em\lower.7ex\hbox{E}\kern-.125emX}}
\begin{document}

%%%%%%%%%%%%%%%%%%%%%%%%%%%%%%%%%%%%%%%%%%%%%%%%%%%%%%%%%%%
%% Titel, Authors, and Thanks
%%%%%%%%%%%%%%%%%%%%%%%%%%%%%%%%%%%%%%%%%%%%%%%%%%%%%%%%%%%
%\title{The Impact of Modeling Mismatches \\ on Near-Field Beamfocusing\\
\title{Physically Consistent Evaluation of\\ Commonly Used Near-Field Models\\
\thanks{The authors would like to thank Dr. Raphael Rolny for discussions and suggestions. This work was funded in part by armasuisse.
Emojis by Twitter, Inc.~and other contributors are licensed under \href{https://creativecommons.org/licenses/by/4.0/}{CC-BY 4.0}.}
}
\author{\IEEEauthorblockN{Georg Schwan, Alexander Stutz-Tirri, and Christoph Studer
\\[0.3cm]
\em ETH Zurich, Switzerland; email: gschwan@ethz.ch, alstutz@ethz.ch, and studer@ethz.ch
}
}

\maketitle

%%%%%%%%%%%%%%%%%%%%%%%%%%%%%%%%%%%%%%%%%%%%%%%%%%%%%%%%%%%
%% Abstract
%%%%%%%%%%%%%%%%%%%%%%%%%%%%%%%%%%%%%%%%%%%%%%%%%%%%%%%%%%%

%%%%%%%%%%%%%%%%%%%%%%%%%%%%%%%%%%%%%%%%%%%%%%%%%%%%%%%%%%%
%% Abstract
%%%%%%%%%%%%%%%%%%%%%%%%%%%%%%%%%%%%%%%%%%%%%%%%%%%%%%%%%%%
\begin{abstract}
Near-field multi-antenna wireless communication has attracted growing research interest in recent years. Despite this development, most of the current literature on antennas and reflecting structures relies on simplified models, whose validity for real systems remains unclear. In this paper, we introduce a physically consistent near-field model, which we use to evaluate commonly used models. Our results indicate that common models are sufficient for basic beamfocusing, but fail to accurately predict the sidelobes and frequency dependence of reflecting structures.
\end{abstract}

% %%%%%%%%%%%%%%%%%%%%%%%%%%%%%%%%%%%%%%%%%%%%%%%%%%%%%%%%%%%
% %% Keywords
% %%%%%%%%%%%%%%%%%%%%%%%%%%%%%%%%%%%%%%%%%%%%%%%%%%%%%%%%%%%
% \begin{IEEEkeywords}  
%     %
%     \as{TBD}
%     %
% \end{IEEEkeywords}

%%%%%%%%%%%%%%%%%%%%%%%%%%%%%%%%%%%%%%%%%%%%%%%%%%%%%%%%%%%
%% Body
%%%%%%%%%%%%%%%%%%%%%%%%%%%%%%%%%%%%%%%%%%%%%%%%%%%%%%%%%%%

%%%%%%%%%%%%%%%%%%%%%%%%%%%%%%%%%%%%%%%%%%%%%%%%%%%%%%%%%%%
%% Introduction
%%%%%%%%%%%%%%%%%%%%%%%%%%%%%%%%%%%%%%%%%%%%%%%%%%%%%%%%%%%
\section{Introduction}\label{sec:introduction}
Near-field communication research has gained significant attention in the signal-processing and communication communities~\cite{special_issue_on_near_field_singal_processing}.
Despite this development, the existing literature relies almost exclusively on simplified models based on non-physically consistent assumptions---this raises the fundamental question of whether such commonly used near-field models are sufficient to represent the behavior of real systems.

\blankfootnote{
{\em Notation:} We denote general vectors and matrices using lowercase (e.g., $\vect{a}$) and uppercase (e.g., $\mat{A}$) boldface, respectively.
For phasors (cf.~\cite[Def.~1]{stutz_schwan_studer_efficient_and_physically_consistent_modeling_of_reconfigurable_electromagnetic_structures}) and phasor vectors, we use pink sans-serif (e.g.,~$\phs{a}$) and pink sans-serif boldface (e.g.,~$\phv{a}$).
We indicate the transpose and conjugate transpose by the superscripts~$^\T$ and $^\He$.
We use~$\diag(\vect{a})$ for the diagonal matrix with the elements of the vector~$\vect{a}$ on the main diagonal. We use~$\|\cdot\|_2$ for the Euclidean norm, and~${|\cdot|}$ for absolute values.
We use blackboard bold font for operators (e.g.,~$\mathbb{S}$) and denote proportionality by $\propto$.
We select the $i$th element of the vector~$\vect{a}$ via~$[\vect{a}]_i$ and denote the element-wise conjugate of the vector~$\vect{z}$ as~$\overline{\vect{z}}$.
At frequency $f$, we define the free-space wavenumber $k\triangleq 2\pi f\sqrt{\mu_\up{0}\varepsilon_\up{0}}$, free-space impedance~$Z_\up{0}\triangleq\sqrt{\mu_\up{0}/\varepsilon_\up{0}}$, and free-space wavelength $\lambda\triangleq (\sqrt{\varepsilon_\up{0} \mu_\up{0}}f)^{-1}$, using the permeability $\mu_\up{0}$ and permittivity $\varepsilon_\up{0}$ of free space.
We adopt the physicist's spherical coordinates~\cite{ISO_quantities_and_units_2_mathematics} $(r,\,\theta,\,\varphi)$ representing radial distance, polar angle, and azimuthal angle, with corresponding local unit vectors $\hat{\vect{r}}$,~$\hat{\vect{\theta}}$, and~$\hat{\vect{\varphi}}$.
Finally, we denote the set of all angles by $\Omega\triangleq[0,\pi]\times[0,2\pi)$.

}

\subsection{Contributions}

We propose a physically consistent\footnote{With \emph{physically consistent}, we refer to the adherence to the laws of classical electromagnetism as described by Maxwell's equations.} sampled near-field model. 
Our model is capable of characterizing general reconfigurable electromagnetic structures (REMS), such as reconfigurable intelligent surfaces (RISs) as well as conventional non-reconfigurable antenna arrays. Concretely, our model predicts the electromagnetic field at a prespecified discrete set of coordinates. The necessary model parameters can be extracted by sampling the electromagnetic field through real-world measurements or via full-wave simulations.\footnote{Full-wave electromagnetic (EM) simulations numerically solve Maxwell's equations. In this work, we use Ansys HFSS~\cite{ansys_hffs}.} The near-field REMS model proposed in this paper builds upon the far-field REMS model originally presented in~\cite{stutz_schwan_studer_efficient_and_physically_consistent_modeling_of_reconfigurable_electromagnetic_structures}.

We leverage our near-field model to conduct a physically consistent evaluation of common near-field models. 
Specifically, we investigate beamfocusing in the following scenarios: (\rom{1})~a uniform linear array (ULA) in free space, (\rom{2})~a ULA with an obstacle positioned between the ULA and the focus point, and (\rom{3})~a reconfigurable intelligent surface (RIS) in free space.
%%%%%%%%%%%%%%%%%%%%%%%%%%%%%%%%%%%%%%%%%%%%%%%%%%%%%%%%%%%
%% Framework
%%%%%%%%%%%%%%%%%%%%%%%%%%%%%%%%%%%%%%%%%%%%%%%%%%%%%%%%%%%

\section{Proposed Sampled Near-Field REMS Model}\label{sec:framework}
\begin{figure*}
    \centering
    \subfloat[Structure of the proposed model.]{
    \centering
    {
    \small
    \begin{tikzpicture}

    \def\boxW{0.7}
    \def\boxH{2.6}

    \def\centRF{0};
    \def\centTN{1.4};
    \def\centRS{2.8};

    \def\centROI{5.0}
    
    \def\wROI{3}
    \def\numPoints{3}

    \def\lineH{0.4}

    \foreach \h in {\lineH, -\lineH} {
        \draw[line width=.6] (\centRF,\h) -- (\centRS,\h);
        
        \foreach \x in {\centRF/2+\centTN/2, \centTN/2+\centRS/2} {
            \draw  [line width=.7pt, fill=white] (\x,\h) ellipse (0.12 and 0.12);
            \draw  [line width=.5pt, fill=black] (\x,\h) ellipse (0.02 and 0.02);
        }
    }
            
    \begin{scope}[shift={(\centRF,0)}]
        \draw [fill=white] (-\boxW/2,-\boxH/2) rectangle (\boxW/2,\boxH/2) node[pos=.5, rotate=90] {RF frontend};
    \end{scope}
    
    \begin{scope}[shift={(\centTN,0)}]
        \draw [fill=white] (-\boxW/2,-\boxH/2) rectangle (\boxW/2,\boxH/2) node[pos=.5, rotate=90] {tuning network};
    \end{scope}
    
    \begin{scope}[shift={(\centRS,0)}]
        \draw [fill=white] (-\boxW/2,-\boxH/2) rectangle (\boxW/2,\boxH/2) node[pos=.5, rotate=90] {radiating structure};
    \end{scope}

    \draw  [line width=.7pt, dashed, color=grey] (\centRF-\boxW/2-0.2,-\boxH/2-0.2) rectangle (\centRS+\boxW/2+0.2,\boxH/2+0.2);

    \node[anchor=south west] at (\centRF-\boxW/2-0.2,\boxH/2+0.2) {\color{grey} REMS};
    
    \foreach \x in {\centRF/2+\centTN/2, \centTN/2+\centRS/2} {
        \begin{scope}[shift={(\x,0)}]
        
            \draw  [line width=.5pt, fill=grey, color=grey] (0,.14)  ellipse (0.02 and 0.02);
            \draw  [line width=.5pt, fill=grey, color=grey] (0,0) ellipse (0.02 and 0.02);
            \draw  [line width=.5pt, fill=grey, color=grey] (0,-0.14) ellipse (0.02 and 0.02);
            
            \begin{scope}[shift={(0,-1.15-\lineH)}]
                \draw[line width=0.3pt, color=black, arrows = {-Stealth[inset=0, length=6pt, angle'=25]}] (0,.2) -- (0,.9);
            \end{scope}
        \end{scope}
    }
    
    \begin{scope}[shift={(0,-1.16-\lineH)}]
    \node[color=black, fill=white] at (\centRF/2+\centTN/2,0) {$N$ ports};
    \end{scope}
    
    \begin{scope}[shift={(0,-1.16-\lineH)}]
    \node[color=black, fill=white] at (\centTN/2+\centRS/2,0) {$M$ ports};
    \end{scope}

    % points

    \foreach \x in {-\numPoints,...,\numPoints} {
        \foreach \y in {-\numPoints,...,\numPoints} {
            \draw  [fill=black] ({\centROI + \x * \wROI / (\numPoints*2+1)}, {\y * \wROI / (\numPoints*2+1)}) ellipse (0.04 and 0.04);
        }
    }

    \node[shift={(\centROI,0)}, anchor=south] at (0, \wROI/2) {\color{grey} region of interest};
    \draw  [shift={(\centROI,0)}, line width=.7pt, dashed, color=grey] (-\wROI/2, -\wROI/2) rectangle (\wROI/2, \wROI/2);

    \begin{scope}[shift={({\centROI - 1 * \wROI / (\numPoints*2+1)}, {-3 * \wROI / (\numPoints*2+1)})}]
    
        \draw [fill=white, color=white] (-0.1,-0.1) rectangle (0.1,0.1);

    \end{scope}
    
    \begin{scope}[shift={(\centROI-0.5,-\wROI/2.)}]

        \begin{scope}[rotate=-42]
            \draw[fill=white, color=white] (-0.6, -0.2) rectangle (0.6, 0.2);
                
            \draw[line width=.8pt] (0.1,-0.2) -- (0.1,0.2);
            \draw[line width=.8pt] (0,-0.2) -- (0,0.2);
            \draw[line width=.8pt] (0.2,-0.2) -- (0.2,0.2);

            \draw [line width=1.5pt, arrows = {-Stealth[inset=0, length=8pt, angle'=35]}] (.6,0) -- (-.6,0);
        \end{scope}

        \draw[fill=white, color=white] (0.7, -0.1) rectangle (1.9, 0.1);
        \node [align=left] at (1.4,-0.2) {\color{black_light} \footnotesize incoming\\ \footnotesize plane wave};
        
        \draw [color=black_light, line width=.8pt] plot[smooth, tension=1.4] coordinates {(0.7,0.05) (0.5,0.1) (0.3,0.1)};
        
    \end{scope}

    \begin{scope}[shift={({\centROI + 1 * \wROI / (\numPoints*2+1)}, {1 * \wROI / (\numPoints*2+1)})}]
    
        \draw [line width=1.pt, color=green, arrows = {-Stealth[inset=0, length=5pt, angle'=35]}] (0,0) -- (0.6,0.15);
            
        \draw [line width=1.pt, color=blue, arrows = {-Stealth[inset=0, length=5pt, angle'=35]}] (0,0) -- (0.15,0.5);

        \node [color=blue] at (-0.15, 0.25) {$E$};
        
        \node [color=green] at (0.6, 0.33) {$H$};
        
        \node [align=left, fill=white] at (0.38,-0.66) {\color{black_light} \footnotesize sample of\\ \footnotesize EM field};
        
        \draw [color=black_light, line width=.8pt] plot[smooth, tension=1.4] coordinates {(0.3,-0.3) (0.2, -0.15) (0.1,-0.1)};

    \end{scope}
    \end{tikzpicture}}
    }
    \hfill
    \subfloat[Example results of the proposed model.]{
    \small
    \begin{tikzpicture}
    
        \node[anchor=south west,inner sep=0] at (0,0) {\includegraphics{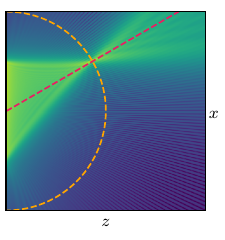}};
        \draw (-0.3,1.38) rectangle (0.11,2.9) node[pos=.5, rotate=90] {\color{grey} REMS};
        
        \node[anchor=south] at (1.75, 3.8) {\color{grey} region of interest};

        \node[anchor=south west,inner sep=0] at (4,0) {\includegraphics{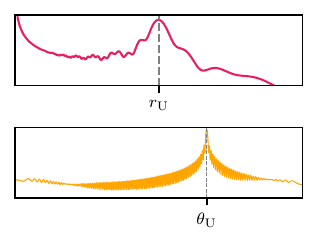}};
        
        \draw [color=black_light, line width=.5pt] plot[smooth, tension=1.4] coordinates {(1.9,1.57)  (3.5, 1.5) (5.1,1.22)};
        
        \draw [color=black_light, line width=.5pt] plot[smooth, tension=1.4] coordinates {(2.4,3.27) (3.7, 2.97) (5.,2.97)};

        \draw[decoration={brace},decorate, color=black!100]
            (0.15,2.25) -- node[anchor=300] {$r_\up{U}$} (1.48,3.04);
        \draw[black!20,->,>=stealth] (0.128,2.18) coordinate (O) -- (1.75,2.18);

        \coordinate (A) at (0.09,2.18);
        
        \draw[black!10,->,>=stealth] ($(A)+(0:1)$) arc(0:30:1) node[midway,anchor=190,circle,] () {$\theta_\up{U}$};

        \node[anchor=south] at (2.7, 2.4) {\color{black!10}focus point};
        \draw[black!10,->,>=stealth] (1.9, 2.85) -- (1.7,3.0);
    \end{tikzpicture}
    \label{fig:example_plots}
    }
    
\caption{(a) Structure of the proposed model: The \emph{RF frontend} comprises $N$ power amplifiers, the \emph{tuning network} represents the reconfigurable RF circuitry, and the \emph{radiating structure} consists of $M$ antenna elements. The inputs to the model are the voltages of the power amplifiers and incoming plane waves; the output is the
electromagnetic field at a prespecified discrete set of coordinates. (b) Beamfocusing example from~\fref{sec:beam_focus_scenario}, showing the sampled energy density $u(\,r,\theta,\varphi)$ in the region of interest.}
\label{fig:rems_structure}
\end{figure*}
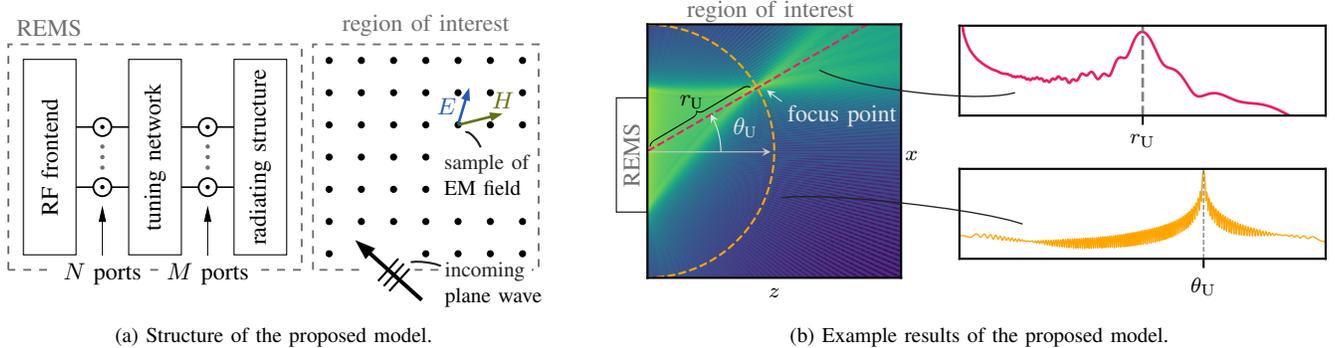

Our proposed physically consistent near-field REMS model is illustrated in~\fref{fig:rems_structure}. For self-consistency, in the following, we restate certain concepts from~\cite[Sec.~II]{stutz_schwan_studer_efficient_and_physically_consistent_modeling_of_reconfigurable_electromagnetic_structures}. 

\setcounter{equation}{9}
\begin{figure*}[bh!]
    \vspace{-0cm}
    \begingroup
    \addtolength{\jot}{-.3em}
    \hrule
    \vspace{-0.2cm}
    \begin{align}
    \label{eq:gain1}
    \mathbb{G}_{\phv{v}_\up{Tx}}^{\phv{a}_\up{N}}
    &\triangleq
    \oper{S}_{\up{R}_\up{NR}}
    \left(
        \mat{I}_{M}
        -\mat{L}_2
     \right)^{-1}
     \mat{S}_{\up{T}_\up{RT}}
     \left(
        \mat{I}_{N}
        -\mat{L}_1
        -\mat{L}_3
    \right)^{-1}
    \mat{K}_{\phv{v}_\up{Tx}}\\
    \mathbb{G}_{\phv{b}_\up{F}}^{\phv{a}_\up{N}}
    &\triangleq
    \oper{S}_{\up{R}_\up{NR}}
    (
        \mat{S}_{\up{T}_\up{RT}}
        \mat{S}_\up{RF}
        \big(
            \mat{I}_{N}
            -\mat{L}_5
        \big)^{-1}
        \mat{S}_{\up{T}_\up{TR}}
        +
        \mat{S}_{\up{T}_\up{RR}}
    )
    (
        \mat{I}_{M}
        -\mat{L}_6
        -\mat{L}_7
    )^{-1}
    \oper{S}_{\up{R}_\up{RF}}
    \,+\,
    \oper{S}_{\up{R}_\up{NF}}
    \label{eq:gain2}
    \end{align}
    \vspace{-0.4cm}
    \hrule
    \vspace{-0.2cm}
    \subfloat{
    \begin{minipage}{0.49\textwidth}
        \begin{align}
        \label{eq:aux1}
        \mat{K}_{\phv{v}_\up{Tx}}
        &\triangleq
        (\mat{Z}_{\up{Tx}}+R_\up{0}\mat{I}_{N})^{-1}\sqrt{R_\up{0}}
        \\
        \mat{S}_{\up{RF}}
        &\triangleq
        (\mat{Z}_{\up{Tx}}+R_\up{0}\mat{I}_{N})^{-1}(\mat{Z}_{\up{Tx}}-R_\up{0}\mat{I}_{N})
        \\
        \mat{L}_\up{1}
        &\triangleq
        \mat{S}_\up{RF}
        \mat{S}_{\up{T}_\up{TT}}
        \\
        \mat{L}_\up{2}
        &\triangleq
        \mat{S}_{\up{T}_\up{RR}}
        \oper{S}_{\up{R}_\up{RR}}  
        \end{align}
    \end{minipage}
    }
    \hfill
    \subfloat{
    \begin{minipage}{0.49\textwidth}
        \begin{align}
        \mat{L}_\up{3}
        &\triangleq
        \mat{S}_\up{RF}
        \mat{S}_{\up{T}_\up{TR}}
        \oper{S}_{\up{R}_\up{RR}}
        \left(
        \mat{I}_{M}
        -\mat{L}_\up{2}
        \right)^{-1}
        \mat{S}_{\up{T}_\up{RT}}
        \\
        \mat{L}_5
        &\triangleq
        \mat{S}_{\up{T}_\up{TT}}
        \mat{S}_\up{RF}
        \\
        \mat{L}_6
        &\triangleq
        \oper{S}_{\up{R}_\up{RR}}
        \mat{S}_{\up{T}_\up{RR}}
        \\
        \mat{L}_7
        &\triangleq
        \oper{S}_{\up{R}_\up{RR}}
        \mat{S}_{\up{T}_\up{RT}}
        \mat{S}_\up{RF}
        \left(
        \mat{I}_{N}
        -\mat{L}_5
        \right)^{-1}
        \mat{S}_{\up{T}_\up{TR}}.
        \label{eq:aux2}
        \end{align}
    \end{minipage}
    }
    \endgroup
\end{figure*}
\setcounter{equation}{0}

\subsection{Sampled Near-Field REMS Model}
We start by describing the external behavior of a REMS.
We denote the electric and magnetic field phasor vectors as $\phv{E}(r,\theta,\varphi)$ and $\phv{H}(r,\theta,\varphi)$, respectively. 
Sufficiently far away from the radiating structure, the electric field can be approximated as in the right-hand side of~\cite[Eq.~5.12]{nieto_vesperinas_scattering_and_diffraction_in_physical_optics}
%Sufficiently far away from the radiating structure, in the limit as the distance approaches infinity, the electric field can be decomposed as
\begin{align}
    \phv{E}(r,\theta,\varphi)
    &\approx
    \phv{E}_\up{FF}^\swarrow(\theta,\varphi)\frac{e^{+jkr}}{r}
    \,+\,
    \phv{E}_\up{FF}^\nearrow(\theta,\varphi)\frac{e^{-jkr}}{r},
    \label{eq:far_field_superposition}
\end{align}
where $\phv{E}_\up{FF}^\swarrow(\theta,\varphi)$ and $\phv{E}_\up{FF}^\nearrow(\theta,\varphi)$ are the incoming and outgoing far-field patterns, which are three-dimensional phasor vectors. 

To formalize the electromagnetic (EM) waves incident from sufficiently far away, we use the \emph{incoming spherical power-wave pattern}, which is defined as~\cite[Def.~3]{stutz_schwan_studer_efficient_and_physically_consistent_modeling_of_reconfigurable_electromagnetic_structures}
\begin{align}
    \phv{b}_\up{F}:\Omega\rightarrow\mathbb{C}^2,
    \,
    (\theta,\varphi)\mapsto
    \frac{1}{\sqrt{Z_0}}
    \begin{bmatrix}
        \hat{\vect{\theta}}^\T\phv{E}_\up{FF}^\swarrow(\theta,\varphi)
        \\
        \hat{\vect{\varphi}}^\T\phv{E}_\up{FF}^\swarrow(\theta,\varphi)
    \end{bmatrix}\!.
    \label{eq:definition_far_field_power_waves_b}
\end{align}
To formalize the outgoing EM field in the region $\overline{\mathcal{V}}$ exterior to the REMS (including the near-field region), we define the \emph{outgoing EM field} vector as
\begin{align}
    \phv{a}_\up{N}:\overline{\mathcal{V}}\rightarrow\mathbb{C}^6,
    \,
    (r,\theta,\varphi)\mapsto
    \begin{bmatrix}
        \frac{1}{\sqrt{Z_0}}\phv{E}^\nearrow(r, \theta,\varphi)
        \\
        \sqrt{Z_0}\phv{H}^\nearrow(r, \theta,\varphi)
    \end{bmatrix},
    \label{eq:definition_far_field_power_waves_a}
\end{align}
where $\phv{E}^\nearrow$ and $\phv{H}^\nearrow$ are the outgoing components of the electromagnetic field, i.e., they represent the fields that are directly radiated from or scattered by the REMS.\footnote{The total EM field is obtained by superposing this outgoing field with the incident field, which is characterized by the incoming spherical power-wave pattern $\phv{b}_\up{F}$.}

To describe the internal behavior of a REMS, we rely on scattering parameters and \emph{circuit-theoretic power-waves} (cf.~\cite[Eq.~1]{kurokawa_power_waves_and_scattering}).
Our model comprises three subsystems (see \fref{fig:rems_structure}): the \emph{RF frontend}, the \emph{radiating structure}, and the \emph{tuning network}.

The \emph{RF frontend} consists of $N\!\in\!\mathbb{Z}_{\geq 0}$ power amplifiers (PAs), modeled by their Thévenin-equivalent circuits. The equivalent voltages and impedances are denoted by 
\begin{align}
        \phv{v}_\up{Tx}
        &\hspace{-1pt}\triangleq
        [\phs{v}_{\up{Tx},1}
        \cdots\,
        \phs{v}_{\up{Tx},N}
        ]^\T
    \\
        \mat{Z}_\up{Tx}
        &\hspace{-1pt}\triangleq
        \diag\big([\mathrm{Z}_{\up{Tx},1}
        \cdots\,
        \mathrm{Z}_{\up{Tx},N}]\big).
\end{align}

The \emph{radiating structure} consists of $M\in\mathbb{N}$ antennas.
Let~$\phv{a}_{\up{R}}\in\mathbb{C}^M$ denote the incoming circuit-theoretic power-waves (from the radiating structure's perspective) on the $M$ antenna ports. The resulting outgoing circuit-theoretic power-waves $\phv{b}_{\up{R}}\in\mathbb{C}^M$ (from the radiating structure's perspective)  and the outgoing EM field $\phv{a}_\up{N}$ are linearly related by%\footnote{Linearity of the operator $\oper{S}_\up{R}$ follows directly from the linearity of Maxwell's equations and the assumption that only linear materials are used.}
\begin{align}
    \begin{bmatrix} \phv{b}_{\up{R}} \\ \phv{a}_\up{N} \end{bmatrix} = 
    \underbrace{
    \begin{bmatrix}
    \oper{S}_{\up{R}_{\up{R}\up{R}}} & \oper{S}_{\up{R}_{\up{R}\up{F}}} \\
    \oper{S}_{\up{R}_{\up{N}\up{R}}} & \oper{S}_{\up{R}_{\up{N}\up{F}}}
    \end{bmatrix}
    }_{\triangleq\,\oper{S}_\up{R}}
    \begin{bmatrix} \phv{a}_{\up{R}} \\ \phv{b}_\up{F} \end{bmatrix}.
    \label{eq:matrix_SF}
\end{align}

The \emph{tuning network} contains the reconfigurable RF circuitry utilized for beamfocusing in the scattering system.\footnote{In the transmitting system, the tuning network can be used for impedance matching and analog beamforming, as demonstrated in~\cite{ferencikova_schwan_joint_beamforming_and_matching}.}
We represent the tuning network as a multiport as follows: 
\begin{align}
    \begin{bmatrix} \phv{b}_\up{T} \\ \phv{a}_\up{R} \end{bmatrix} = 
    \begin{bmatrix}
    \mat{S}_{\up{T}_{\up{T}\up{T}}} & \mat{S}_{\up{T}_{\up{T}\up{R}}} \\
    \mat{S}_{\up{T}_{\up{R}\up{T}}} & \mat{S}_{\up{T}_{\up{R}\up{R}}}
    \end{bmatrix}
    \begin{bmatrix} \phv{a}_\up{T} \\ \phv{b}_\up{R} \end{bmatrix}\!.
    \label{eq:matrix_ST}
\end{align}
Here, $\phv{a}_\up{T}\in\mathbb{C}^{N}$ and $\phv{b}_\up{T}\in\mathbb{C}^{N}$ represent the incoming and outgoing circuit-theoretic power-waves (from the tuning network’s perspective) at the ports connecting the tuning network to the RF frontend.

\subsection{Input-Output Relationship}
The system inputs consist of the PA equivalent voltages~$\phv{v}_\up{Tx}$ and the incoming spherical power-wave pattern $\phv{b}_\up{F}$. The system output is given by the outgoing EM field vector $\phv{a}_\up{N}$. 
The resulting input-output relation can be expressed as follows:
\begin{align}
    \phv{a}_\up{N}
    &=
    \mathmakebox[\widthof{$\mathbb{G}_{\phv{v}_\up{Tx}}^{\phv{v}_\up{Rx}}\phv{v}_\up{Tx}$}][c]{\mathbb{G}_{\phv{v}_\up{Tx}}^{\phv{a}_\up{N}}\phv{v}_\up{Tx}}
    +
    \mathmakebox[\widthof{$\mathbb{G}_{\phv{b}_\up{F}}^{\phv{v}_\up{Rx}}\phv{b}_\up{F}$}][c]{\mathbb{G}_{\phv{b}_\up{F}}^{\phv{a}_\up{N}}\phv{b}_\up{F}}.
\end{align}
The respective gain operators are defined in~\fref{eq:gain1} and~\fref{eq:gain2} using the auxiliary matrices defined in~\fref{eq:aux1} to~\fref{eq:aux2}.\footnote{$R_\up{0}$ denotes the reference impedance of the scattering parameters (\SI{50}{\ohm}).}

\subsection{Energy Density}
The \emph{energy density} of the electromagnetic field at a specific coordinate $(r,\theta,\varphi)$ is given by~\cite[Eq.~8.13]{griffiths_introduction_to_electrodynamics}\footnote{When phasors are formulated using a peak-value convention, an additional division factor of two is included in this formula. However, our phasor definition utilizes the root-mean-square (RMS) value convention, which obviates this factor.}
\begin{align}
    \!u(\,r,\theta,\varphi)
    \triangleq
    \frac{1}{2}(\varepsilon_\up{0}\|\phv{E}\|_2^2 + \mu_\up{0}\|\phv{H}\|_2^2)
    =
    \frac{\sqrt{\mu_\up{0}\varepsilon_\up{0}}}{2} \|\phv{a}_\up{N}(r, \theta,\varphi)\|_{2}^2.
\end{align}
\setcounter{equation}{19}

%%%%%%%%%%%%%%%%%%%%%%%%%%%%%%%%%%%%%%%%%%%%%%%%%%%%%%%%%%%
%% Results
%%%%%%%%%%%%%%%%%%%%%%%%%%%%%%%%%%%%%%%%%%%%%%%%%%%%%%%%%%%

\section{Test Scenarios}\label{sec:results}

We now use our near-field model from~\fref{sec:framework} to evaluate commonly used near-field models across three scenarios. \fref{tab:overview} provides an overview of these scenarios and summarizes the corresponding results. 
In each scenario, the region of interest is located in the $x$-$z$ plane in front of the REMS.
The parameters for our physically consistent model, specifically for $\oper{S}_\up{R}$, were extracted using full-wave EM simulations in Ansys HFSS.

\subsection{Scenario~\rom{1}}\label{sec:beam_focus_scenario}
\subsubsection{Setup}
\begin{figure}[t]
    \centering
    \subfloat[Patch antenna]{
        \centering
        {
        \small
        \includegraphics[width=0.41\linewidth]{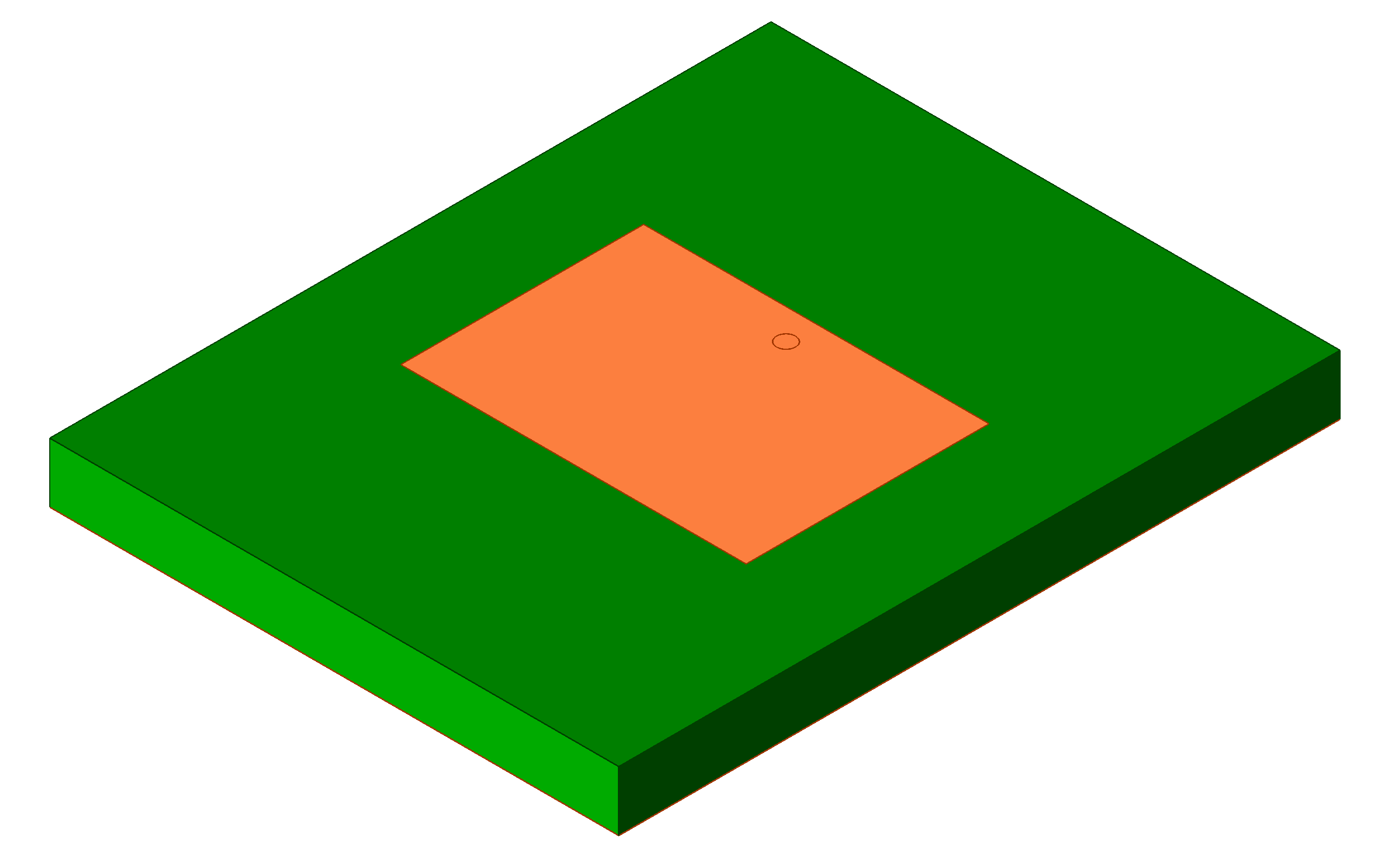}
        }
        \label{fig:hfss_patch}
    }
    \subfloat[RIS unit cell]{
        \centering
        {
        \small
        \includegraphics[width=0.41\linewidth]{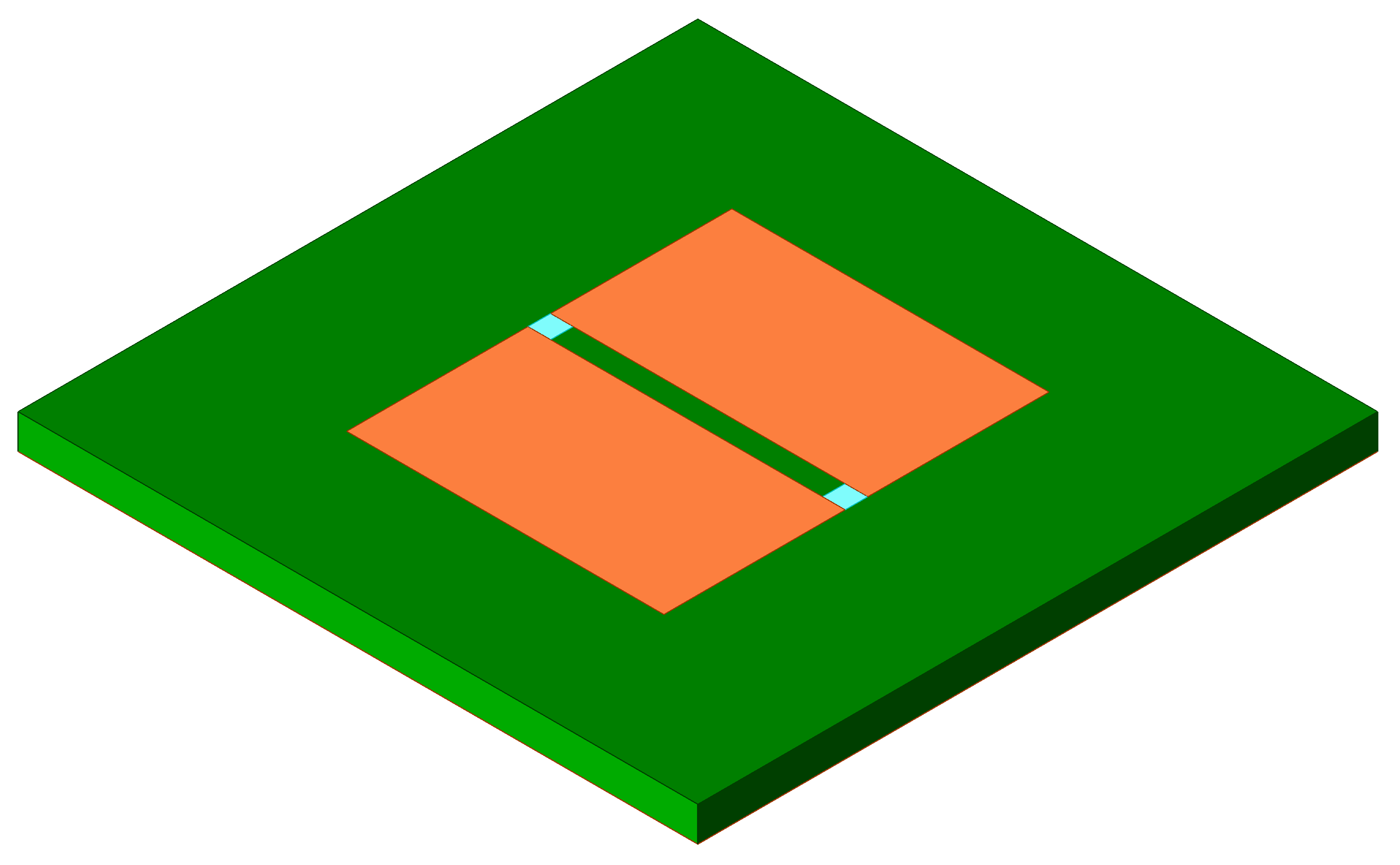}
        }
        \label{fig:hfss_unit_cell}
    }
\caption{Ansys HFSS screenshots of a patch-antenna element and a RIS unit cell. The ULA elements are excited via a coaxial feed (not visible); the RIS unit cells feature two ports (in light blue) instead of the varactor diodes.}

\label{fig:hfss}
\end{figure}

We consider a uniform linear array (ULA) transmitter operating at \SI{10}{\giga \hertz}. The array employs 128 patch antennas (see~\fref{fig:hfss_patch}) with $\lambda/2$ inter-antenna spacing.

The objective is to focus a beam onto a specific focus point $(r_\up{U}, \theta_\up{U},\varphi_\up{U}) = (\SI{1.92}{\meter}, 30^\circ,0^\circ)$ located within the near-field region. Concretely, we maximize the energy density $u(r_\up{U}, \theta_\up{U},\varphi_\up{U})$ at the focus point.\footnotemark

% Please add the following required packages to your document preamble:
% \usepackage{multirow}
\begin{table}[]
\caption{Summary of Test Scenarios and Results}
\label{tab:overview}
\renewcommand{\arraystretch}{1.3}
\resizebox{\columnwidth}{!}{
\begin{tabular}{|cll|l|c|c|}
\hline
\multicolumn{3}{|c|}{\textbf{Scenario}} &
  \textbf{Model under Test} &
  \textbf{Fig.} &
  \textbf{Works?} \\ \hline  \hline
\multicolumn{1}{|c|}{\multirow{3}{*}{\rom{1}}} &
  \multicolumn{1}{c|}{\multirow{3}{*}{ULA}} &
  beamfocusing &
  \multirow{3}{*}{spherical-wave model} &
  \ref{fig:beamfocus_line}, \ref{fig:beamfocus_circle}, \ref{fig:example_plots} &
  \Large \raisebox{-0.73mm}{\texttwemoji{beaming face with smiling eyes}} \\ \cline{3-3} \cline{5-6} 
\multicolumn{1}{|c|}{} &
  \multicolumn{1}{l|}{} &
  tapering &
   &
  \ref{fig:tapering_line}, \ref{fig:tapering_circle}, \ref{fig:tapering_rect} &
  \Large\raisebox{-0.82mm}{\texttwemoji{beaming face with smiling eyes}} \\ \cline{3-3} \cline{5-6} 
\multicolumn{1}{|c|}{} &
  \multicolumn{1}{l|}{} &
  frequency dependence &
   &
  \ref{fig:freq_line} &
  \Large\raisebox{-0.82mm}{\texttwemoji{expressionless face}} \\ \hline
\multicolumn{1}{|c|}{\rom{2}} &
  \multicolumn{1}{c|}{ULA} &
  \begin{tabular}[c]{@{}l@{}}beamfocusing behind \\ an obstacle\end{tabular} &
  \begin{tabular}[c]{@{}l@{}}spherical-wave model\\ with perfect absorber\end{tabular} &
  \ref{fig:head_line}, \ref{fig:head_rect} &
  \Large\raisebox{-0.82mm}{\texttwemoji{expressionless face}} \\ \hline
\multicolumn{1}{|c|}{\multirow{2}{*}{\rom{3}}} &
  \multicolumn{1}{c|}{\multirow{2}{*}{RIS}} &
  beamfocusing &
  \multirow{2}{*}{spherical-wave model} &
  \ref{fig:ris_line}, \ref{fig:ris_circle}, \ref{fig:ris_rect} &
  \Large\raisebox{-0.73mm}{\texttwemoji{expressionless face}} \\ \cline{3-3} \cline{5-6} 
\multicolumn{1}{|c|}{} &
  \multicolumn{1}{l|}{} &
  frequency dependence &
   &
  \ref{fig:ris_freq} &
  \Large\raisebox{-0.82mm}{\texttwemoji{worried face}} \\ \hline
\end{tabular}

}

\end{table}

\subsubsection{Model under Test}\label{sec:beam_focus_model_under_test}
While common near-field models typically characterize the end-to-end channel gain, we interpret them here as representing energy density. To ensure a fair comparison with our physically consistent model, we normalize both approaches---thus, specifying the energy density up to a constant factor is sufficient. The respective energy density given by the commonly used spherical-wave-based model is as follows:~\cite{yuanwei_near_field_communication_a_tutorial_review}
\begin{align}
    u(r,\theta,\varphi)
    \propto
    |\vect{c}(r, \theta,\varphi)^\T \phv{v}_\up{Tx}|^2.
\end{align}
Here, $\vect{c}(r, \theta,\varphi) \in\mathbb{C}^{N}$ is the array factor, defined as
\begin{align}
    \vect{c}(r, \theta,\varphi)
    \triangleq
    \begin{bmatrix}
    \frac{e^{-jkd_1(r, \theta,\varphi)}}{d_1(r, \theta,\varphi)} & \dots & \frac{e^{-jkd_N(r, \theta,\varphi)}}{d_N(r, \theta,\varphi)}
    \end{bmatrix}^\T,
\label{eq:array_factor}
\end{align}
where $d_i(r, \theta,\varphi)$ denotes the distance between the coordinate~$(r, \theta,\varphi)$ and the $i$th antenna element.

For this model, the optimal beamfocusing vector, in the sense of maximizing the energy density, is given by~\cite{yuanwei_near_field_communication_a_tutorial_review}
\begin{align}
 \phv{v}^\up{SW}_\up{Tx} \triangleq \frac{\overline{\vect{c}}(r_\up{U}, \theta_\up{U},\varphi_\up{U})}{\|\vect{c}(r_\up{U}, \theta_\up{U},\varphi_\up{U})\|_2}.
\end{align}

For our physically consistent model, the optimal beamfocusing vector is given by the optimization problem
\begin{align}
    \phv{v}^\up{PC}_\up{Tx} \in &\argmax_{\|\phv{v}_\up{Tx}\|_{2}^2 = 1} \frac{\sqrt{\mu_\up{0}\varepsilon_\up{0}}}{2} \|\phv{a}_\up{N}(r_\up{U}, \theta_\up{U},\varphi_\up{U})\|_{2}^2,
\end{align}
where the solution corresponds to an eigenvector associated with the largest eigenvalue of $(\mathbb{G}_{\phv{v}_\up{Tx}}^{\phv{a}_\up{N}})^\He\mathbb{G}_{\phv{v}_\up{Tx}}^{\phv{a}_\up{N}}$~\cite[Sec.~III]{ferencikova_schwan_joint_beamforming_and_matching}.

As previously mentioned, we normalize the energy densities to facilitate a comparison between the two models. This normalization is defined relative to the value obtained at the focus point using the focus vector~$\phv{v}^\up{PC}_\up{Tx}$.
Thus, for the commonly used spherical-wave model and our physically consistent model, we define the normalized energy densities as
\begin{align}
 u^\up{SW}(\phv{v}_\up{Tx};\,r,\theta,\varphi)
 &\triangleq
 \frac{|\vect{c}(r, \theta,\varphi)^\T \phv{v}_\up{Tx}|^2}{|\vect{c}(r_\up{U}, \theta_\up{U},\varphi_\up{U})^\T \phv{v}^\up{PC}_\up{Tx}|^2} \\
 u^\up{PC}(\phv{v}_\up{Tx};\,r,\theta,\varphi)
 &\triangleq
 \frac{\|\mathbb{G}_{\phv{v}_\up{Tx}}^{\phv{a}_\up{N}}(r, \theta,\varphi) \phv{v}_\up{Tx}\|_2^2}{\|\mathbb{G}_{\phv{v}_\up{Tx}}^{\phv{a}_\up{N}}(r_\up{U}, \theta_\up{U},\varphi_\up{U}) \phv{v}^\up{PC}_\up{Tx}\|_2^2}.
\end{align}

\subsubsection{Results}
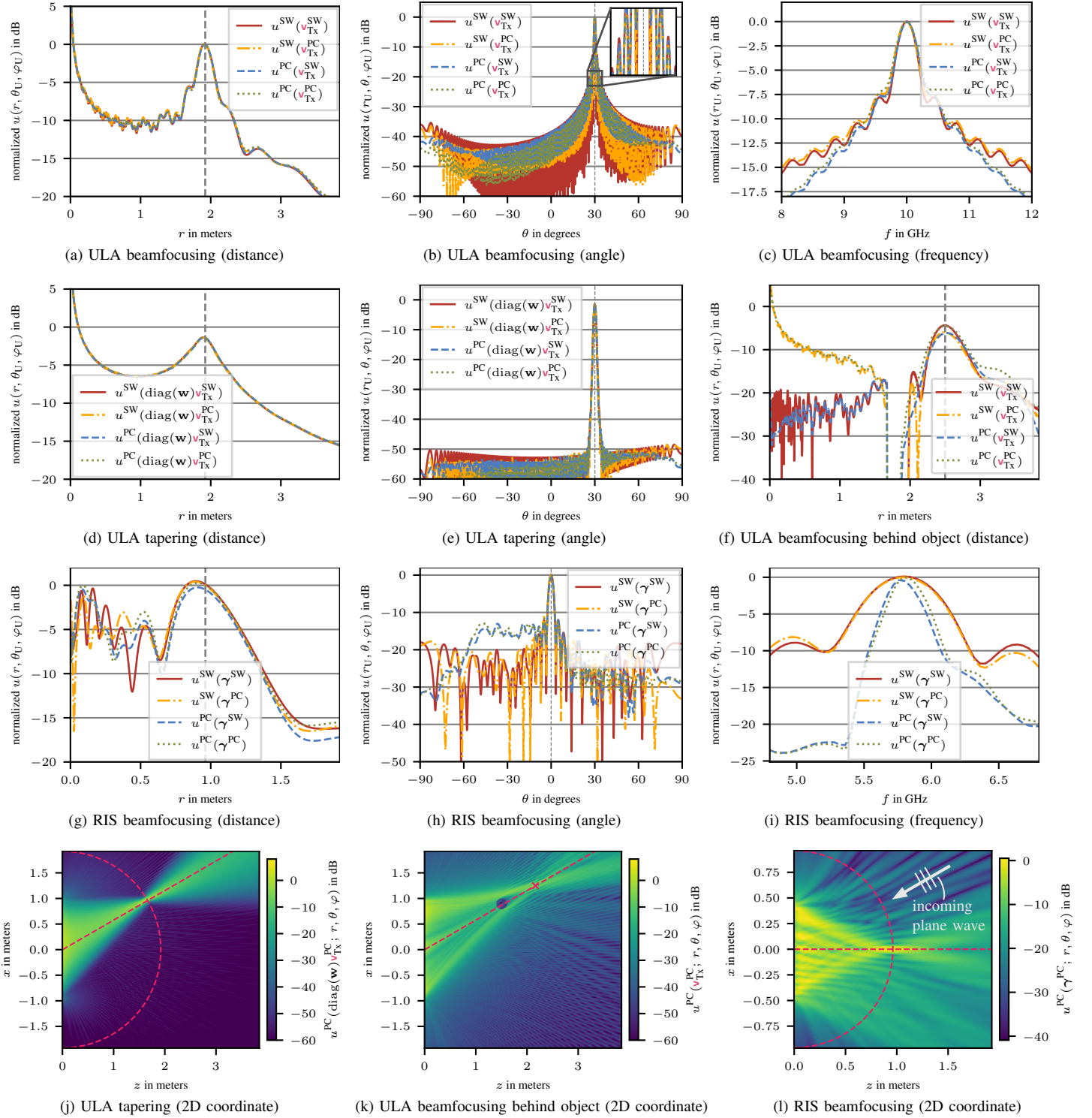
\begin{figure*}[tbp]
    \hspace{-3mm}
    \subfloat[ULA beamfocusing (distance)]{
        {
        \small
        \includegraphics{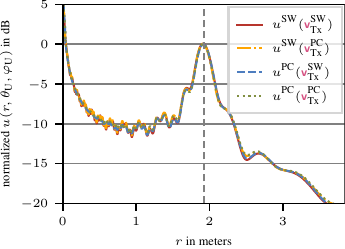}
        }
        \label{fig:beamfocus_line}
    }
    \hspace{-2mm}
    \subfloat[ULA beamfocusing (angle)]{
        {
        \small
        \includegraphics{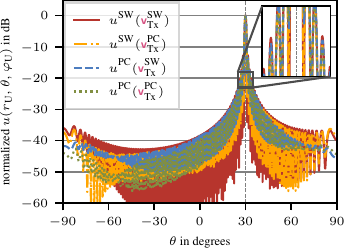}
        }
        \label{fig:beamfocus_circle}
    }
    \hspace{-2mm}
    \subfloat[ULA beamfocusing (frequency)]{
        {
        \small
        \includegraphics{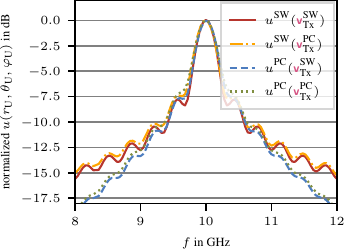}
        }
        \label{fig:freq_line}
    }
    
    \hspace{-3mm}
    \subfloat[ULA tapering (distance)]{
        {
        \small
        \includegraphics{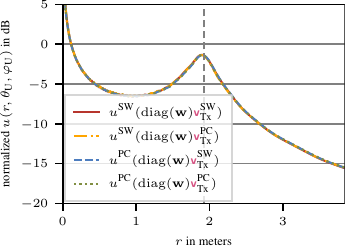}
        }
        \label{fig:tapering_line}
    }
    \hspace{-2mm}
    \subfloat[ULA tapering (angle)]{
        {
        \small
        \includegraphics{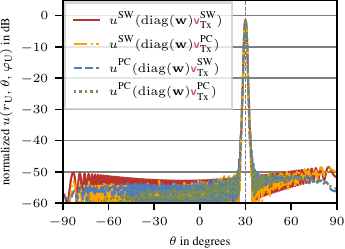}
        }
        \label{fig:tapering_circle}
    }
    \hspace{-2mm}
    \subfloat[ULA beamfocusing behind object (distance)]{
        {
        \small
        \includegraphics{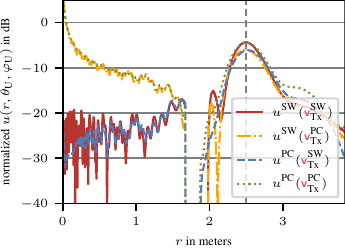}
        }
        \label{fig:head_line}
    }
    
    \hspace{-3mm}
    \subfloat[RIS beamfocusing (distance)]{
        {
        \small
        \includegraphics{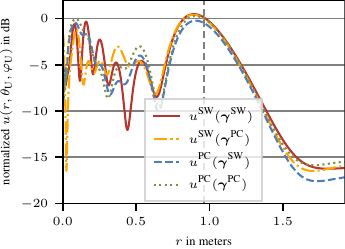}
        }
        \label{fig:ris_line}
    }
    \hspace{-2mm}
    \subfloat[RIS beamfocusing (angle)]{
        {
        \small
        \includegraphics{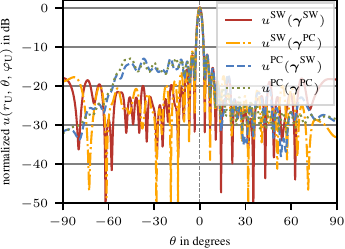}
        }
        \label{fig:ris_circle}
    }
    \hspace{-2mm}
    \subfloat[RIS beamfocusing (frequency)]{
        {
        \small
        \includegraphics{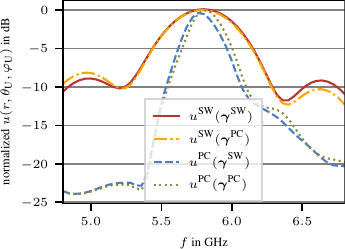}
        }
        \label{fig:ris_freq}
    }
    
    \hspace{-5mm}
    \subfloat[ULA tapering (2D coordinate)]{
        {
        \small
        \includegraphics{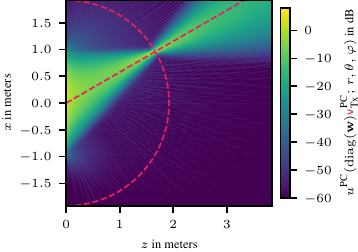}
        }
        \label{fig:tapering_rect}
    }
    \hspace{-2mm}
    \subfloat[ULA beamfocusing behind object (2D coordinate)]{
        {
        \small
        \includegraphics{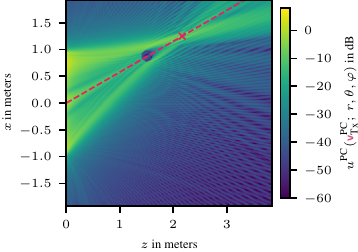}
        }
        \label{fig:head_rect}
    }
    \hspace{-2mm}
    \subfloat[RIS beamfocusing (2D coordinate)]{
        {
        \small
        \begin{tikzpicture}
        
            \node[anchor=south west,inner sep=0] at (0,0) {\includegraphics{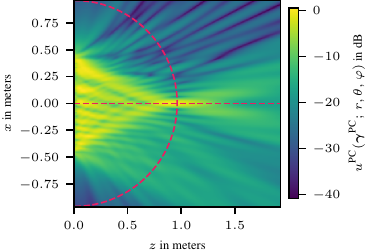}};

            \begin{scope}[shift={(3.5,3.7)}]

                \begin{scope}[rotate=30]
    
                    \draw[line width=.8pt, color=black!10] (0.1,-0.2) -- (0.1,0.2);
                    \draw[line width=.8pt, color=black!10] (0,-0.2) -- (0,0.2);
                    \draw[line width=.8pt, color=black!10] (0.2,-0.2) -- (0.2,0.2);

                    \draw [line width=1.5pt, color=black!10, arrows = {-Stealth[inset=0, length=8pt, angle'=35]}] (.6,0) -- (-.6,0);
                \end{scope}

                \node [align=left] at (.5,-.6) {\color{black!10} \footnotesize incoming\\ \color{black!10}\footnotesize plane wave};
                
                \draw [color=black!10, line width=.8pt] plot[smooth, tension=1.4] coordinates {(0.4,-0.3) (0.4,-0.1) (0.3,0.1)};
                
            \end{scope}
        \end{tikzpicture}
        }
        
        \label{fig:ris_rect}
    }
\caption{Results for Scenarios~\rom{1},~\rom{2}, and~\rom{3} as summarized in \fref{tab:overview}. The figures compare the energy density prediction of a commonly used spherical-wave-based model with the prediction of our physically consistent model. The objective in each scenario is to focus the energy density towards a specific focus point. In Scenario~\rom{1}, the spherical-wave model shows good agreement in the beamforming and tapering cases. However, across different frequencies, deviations increase as the offset from the center frequency grows. In Scenario~\rom{2}, the spherical-wave model is able to focus at the target behind the obstacle. However, the predictive accuracy degrades in the region behind the obstacle. In contrast, by employing our proposed physically consistent model, we achieve superior beamfocusing performance. In Scenario~\rom{3}, while the spherical-wave model focuses energy on the correct target coordinate, it fails to accurately predict the surrounding sidelobes. Furthermore, the commonly used spherical-wave model does not account for specular reflections from the RIS, and, if the frequency deviates from the center frequency, the spherical-wave model is unable to accurately predict the energy density.} 

\label{fig:hfss}
\end{figure*}

\footnotetext{We use the energy density since it is an established physical quantity that serves as a good proxy for the potential power that can be received by electrically small but otherwise general receive antennas. Moreover, if we placed receive antennas into the simulation, they would influence the EM field in an \emph{antenna-specific} way, compromising the generality of our results.}
\fref{fig:beamfocus_line} and \fref{fig:beamfocus_circle} present results for both the spherical-wave model and our model using the focus vectors $\phv{v}^\up{PC}_\up{Tx}$ and $\phv{v}^\up{SW}_\up{Tx}$.\footnote{\fref{fig:example_plots} displays $u^\up{PC}(\phv{v}^\up{PC}_\up{Tx};\,r,\theta,\varphi)$ for this scenario.} We observe strong agreement between the two models, indicating that the commonly used spherical-wave model is sufficient for accurate beamfocusing in this scenario.

\fref{fig:tapering_line}, \fref{fig:tapering_circle}, and \fref{fig:tapering_rect} present the results with tapering applied. Tapering is often employed to reduce sidelobes for beamsteering. To achieve this, the excitation voltages are multiplied by a window $\vect{w} \in\mathbb{R}^{N}$. In this scenario, we select~$\vect{w}$ as a Dolph-Chebyshev window~\cite{dolph_optimze_beam_width_and_side_lobe_level} with a sidelobe level of~\SI{50}{\decibel}. Once again, we observe strong agreement between our physically consistent model and the spherical-wave model.

\fref{fig:freq_line} depicts the normalized energy density at the focus point $(r_\up{U}, \theta_\up{U},\varphi_\up{U})$ across different frequencies. While the spherical-wave model shows good agreement near the center frequency, the discrepancy increases as the frequency deviates from \SI{10}{\giga \hertz}. One cause for this discrepancy is that the spherical-wave model does not account for the narrowband frequency response inherent to the used patch antennas.

\subsection{Scenario~\rom{2}}
\subsubsection{Setup}
We extend the setup from~\fref{sec:beam_focus_scenario} by introducing an obstacle positioned in front of the focus point. The obstacle is modeled as a ball with a radius of \SI{10}{\cm}, located at $(\SI{1.77}{\meter}, 30^\circ, 0^\circ)$. 
We assign the ``humanaverage'' material properties from the HFSS library to the ball. 
In this scenario, the focus point is behind the obstacle at~$(\SI{2.5}{\meter}, 30^\circ, 0^\circ)$.

\subsubsection{Model under Test}
We again evaluate the spherical-wave model from~\fref{sec:beam_focus_model_under_test}, treating the obstacle as a perfect absorber that fully blocks intersecting line-of-sight paths. The optimal beamfocusing vectors $\phv{v}^\up{SW}_\up{Tx}$ and $\phv{v}^\up{PC}_\up{Tx}$ are defined as in~\fref{sec:beam_focus_scenario} using the updated models. Similarly, normalized energy densities follow previous definitions but retain the prior normalization constant to ensure a fair comparison.

\subsubsection{Results}
\fref{fig:head_line} presents the evaluation of both models using both optimal beamfocusing vectors. We observe that the spherical-wave model is sufficient to focus energy density behind the obstacle. However, the predictive accuracy of the spherical-wave model deteriorates in the region behind the obstacle. In contrast, by employing our physically consistent model, we achieve superior beamfocusing performance.

\subsection{Scenario~\rom{3}}

\subsubsection{Setup}
We use a $32 \times 4$ element RIS at \SI{5.8}{\giga \hertz} based on~\cite{okoniewski_realizing_an_tunable_reflectarray_using_varactor_diodes}, with a unit cell spacing of $d = 0.58\lambda$. As shown in~\fref{fig:hfss_unit_cell}, each cell uses two varactor diodes for beamfocusing. The incoming signal is a plane wave incident from~$(\theta_\up{P},\varphi_\up{P}) \triangleq (30^\circ,0^\circ)$. Consistent with the previous scenarios, we maximize the energy density $u(r_\up{U}, \theta_\up{U}, \varphi_\up{U})$ at the focus point $(r_\up{U}, \theta_\up{U}, \varphi_\up{U}) = (\SI{0.96}{\meter},0^\circ,0^\circ)$.

\subsubsection{Model under Test}
For the reason explained in~\fref{sec:beam_focus_model_under_test}, the evaluated commonly used spherical-wave-based near-field model represents the energy density up to a constant factor, and is given as\footnote{For this scenario, the element radiation pattern is assumed to be isotropic.}~\cite{wankai_ris_path_loss_modeling}
\begin{align}
    u(\vect{\gamma};\,r,\theta,\varphi)
    \propto
    |\vect{c}(r, \theta,\varphi)^\T \diag(\vect{\gamma}) \vect{g}(\theta_\up{P},\varphi_\up{P})|^2\!,
\end{align}
where $\vect{\gamma} \in \mathbb{C}^{128}$ denotes the reflection coefficients of the unit cells, and $\vect{g} \in \mathbb{C}^{128}$ represents the channel from the plane wave source to the RIS, defined as
\begin{align}
    \vect{g}(\theta,\varphi)
    \triangleq
    \begin{bmatrix} 
    e^{-jkd(1v_x + 1v_y)} & \dots & e^{-jkd(32v_x + 4v_y)}
    \end{bmatrix}^\T\!,
\end{align}
with $v_x \triangleq \sin(\theta)\cos(\varphi)$ and $v_y \triangleq \sin(\theta)\sin(\varphi)$.

For this model, the optimal reflection coefficient vector is given by~\cite{wankai_ris_path_loss_modeling} 
\begin{align}
    [\vect{\gamma}^\up{SW}]_i
    \triangleq
    \frac{[\overline{\vect{c}}(r_\up{U}, \theta_\up{U},\varphi_\up{U})]_i  [\overline{\vect{g}}(\theta_\up{P},\varphi_\up{P})]_i}{|[\vect{c}(r_\up{U}, \theta_\up{U},\varphi_\up{U})]_i [\vect{g}(\theta_\up{P},\varphi_\up{P})]_i|}.
\end{align}

Determining the optimal configuration $\vect{\gamma}^\up{PC}$ for our physically consistent model is non-trivial due to mutual coupling. Consequently, we compute $\vect{\gamma}^\up{PC}$ by initializing the system with $\vect{\gamma}^\up{SW}$ and performing gradient ascent until convergence is achieved.

We again normalize the energy densities to facilitate a comparison between the two models.
This normalization is defined relative to the value obtained at the focus point using the physically consistent configuration $\vect{\gamma}^\up{PC}$.
Consequently, for the spherical-wave model and our physically consistent model, the normalized energy densities are defined as\footnote{The mapping from the reflection coefficient to the actual impedance of the varactor diode for our physically consistent model is determined via simulations using periodic boundary conditions on the unit cell, following standard practice~\cite{nayeri_reflectarray_antennas}.}
\begin{align}
    u^\up{SW}(\vect{\gamma};\,r,\theta,\varphi)
    &\triangleq
    \frac{|\vect{c}(r, \theta,\varphi)^\T \diag(\vect{\gamma}) \vect{g}(\theta_\up{P},\varphi_\up{P})|^2}{|\vect{c}(r_\up{U}, \theta_\up{U},\varphi_\up{U})^\T \diag(\vect{\gamma}^\up{PC}) \vect{g}(\theta_\up{P},\varphi_\up{P})|^2}
    \raisetag{30pt}
    \\
    u^\up{PC}(\vect{\gamma};\,r,\theta,\varphi)
    &\triangleq
    \frac{\|\mathbb{G}_{\phv{b}_\up{F}}^{\phv{a}_\up{N}}(r, \theta,\varphi) \phv{b}_\up{F}\|_2^2}{\|\mathbb{G}_{\phv{b}_\up{F}}^{\phv{a}_\up{N}}(r_\up{U}, \theta_\up{U},\varphi_\up{U}) \phv{b}_\up{F}\|_2^2}.
\end{align}

\subsubsection{Results}
\fref{fig:ris_line} and \fref{fig:ris_circle} present the evaluation of both our physically consistent model and the commonly used spherical-wave-based model using the optimized reflection coefficient vectors $\vect{\gamma}^\up{PC}$ and $\vect{\gamma}^\up{SW}$. We observe that the spherical-wave vector $\vect{\gamma}^\up{SW}$ focuses the energy density onto the focus point sufficiently well. However, the spherical-wave model fails to accurately predict the sidelobes surrounding the focus point. A major limitation is that the spherical-wave model does not account for specular reflections.\footnote{Notably, the specular reflection of an RIS is generally uncontrollable. Furthermore, specular reflections influence not only the near-field region but also the far-field region of the RIS.} In contrast, our physically consistent model captures this effect, as~\fref{fig:ris_rect} illustrates.

\fref{fig:ris_freq} depicts the normalized energy density at the focus point $(r_\up{U}, \theta_\up{U},\varphi_\up{U})$ across different frequencies. While the spherical-wave model demonstrates good agreement near the center frequency, its prediction accuracy diminishes as the frequency deviates from \SI{5.8}{\giga \hertz}. This inaccuracy arises because the spherical-wave model ignores the narrowband frequency response inherent to the RIS unit cell.
%%%%%%%%%%%%%%%%%%%%%%%%%%%%%%%%%%%%%%%%%%%%%%%%%%%%%%%%%%%
%% Conclusions
%%%%%%%%%%%%%%%%%%%%%%%%%%%%%%%%%%%%%%%%%%%%%%%%%%%%%%%%%%%
\section{Conclusions}\label{sec:conclusions}
In this work, we have proposed a physically consistent sampled near-field model applicable to general reconfigurable electromagnetic structures (REMS). Our model predicts the electromagnetic field at a prespecified discrete set of coordinates. We then used this model to evaluate commonly used spherical-wave-based near-field models.

\fref{tab:overview} summarizes our results: 
While common models are sufficient for basic beamfocusing, models for reconfigurable intelligent surfaces (RISs) fail to accurately predict sidelobes and frequency dependence. Notably, common RIS models neglect specular reflections, leading to inaccuracies in predicting the overall energy density. 

Our framework demonstrates that accurately modeling the near-field region at a discrete set of coordinates is possible without requiring expert knowledge of antenna design or electromagnetic field theory. For reproducibility, our code and models are available online at \href{https://github.com/IIP-Group/evaluation-of-near-field-models}{https://github.com/IIP-Group/evaluation-of-near-field-models}.

%%%%%%%%%%%%%%%%%%%%%%%%%%%%%%%%%%%%%%%%%%%%%%%%%%%%%%%%%%%
%% Bibliography
%%%%%%%%%%%%%%%%%%%%%%%%%%%%%%%%%%%%%%%%%%%%%%%%%%%%%%%%%%%
\balance
\bstctlcite{IEEEexample:BSTcontrol} % keep repeated author names in the biography
\bibliographystyle{IEEEtran}
\bibliography{bib/publishers,bib/journals_proceedings_ect,bib/library}
\balance

\end{document}